\documentclass[aps,pre,twocolumn,superscriptaddress]{revtex4-1}

\usepackage{amsmath,amsfonts,amssymb,amsthm}
\usepackage{mathtools}
\usepackage{xcolor}
\usepackage[colorlinks=true,linkcolor=black,citecolor=black,urlcolor=black,bookmarks,breaklinks=true]{hyperref}

\usepackage{physics, bm}
\usepackage{siunitx}

\usepackage[export]{adjustbox}

\renewcommand{\vec}[1]{\bm{#1}}

\begin{document} 

\title{Critical pore radius and transport properties\\ of disordered hard- and overlapping-sphere models}

\author{Michael A. Klatt}
\email[]{Email: mklatt@princeton.edu}
\affiliation{Department of Physics, Princeton University, Princeton, New Jersey 08544, USA}
\affiliation{Institut für Theoretische Physik, FAU Erlangen-Nürnberg, Staudtstr. 7, 91058 Erlangen, Germany}
\author{Robert M. Ziff}
\email[]{Email: rziff@umich.edu}
\affiliation{Center for the Study of Complex Systems and Department of Chemical Engineering, University of Michigan, Ann Arbor, Michigan 48109, USA}
\author{Salvatore Torquato}
\email[]{Email: torquato@princeton.edu}
\affiliation{Department of Physics, Princeton University, Princeton, NJ 08544, USA}
\affiliation{Department of Chemistry, Princeton Institute for the Science and Technology of Materials, and Program in Applied and Computational Mathematics, Princeton University, Princeton, New Jersey 08544, USA}
\date{\today}

\begin{abstract} \noindent 
Transport properties of porous media are intimately linked to their 
pore-space microstructures.
We quantify geometrical and topological descriptors of the pore space of 
certain disordered and ordered distributions of spheres, including 
pore-size functions and the critical pore radius $\delta_c$.
We focus on models of porous media derived from maximally random jammed 
sphere packings, overlapping spheres, equilibrium hard spheres, 
``quantizer'' sphere packings, and crystalline sphere packings.
For precise estimates of the percolation thresholds, we use a strict 
relation of the void percolation around sphere configurations to 
weighted bond percolation on the corresponding Voronoi networks.
We use the Newman-Ziff algorithm to determine the percolation threshold 
using universal properties of the cluster size distribution.
The critical pore radius $\delta_c$ is often used as the key 
characteristic length scale that determines the fluid permeability $k$.
A recent study [Torquato.~Adv.~Wat.~Resour.~140,103565 (2020)]
suggested for porous media with a well-connected pore space an 
alternative estimate of $k$ based on the second moment of the pore size 
$\langle \delta^2 \rangle$, which is easier to determine than 
$\delta_c$.
Here, we compare $\delta_c$ to the second moment of the pore size 
$\langle \delta^2 \rangle$, and indeed confirm that, for all porosities 
and all models considered, $\delta_c^2$ is to a good approximation 
proportional to $\langle \delta^2 \rangle$.
However, unlike $\langle \delta^2 \rangle$, the permeability  estimate 
based on $\delta_c^2$ does not predict the correct ranking of $k$ for 
our models.
Thus, we confirm $\langle \delta^2 \rangle$ to be a promising candidate 
for convenient and reliable estimates of the fluid permeability for 
porous media with a well-connected pore space.
Moreover, we compare the fluid permeability of our models with varying 
degrees of order, as measured by the $\tau$ order metric.
We find that (effectively) hyperuniform models tend to have lower values 
of $k$ than their nonhyperuniform counterparts.
Our findings could facilitate the design of porous media with  desirable 
transport properties  via targeted pore statistics.
\end{abstract}

\keywords{Percolation, centroidal Voronoi tessellations, overlapping 
spheres, equilibrium hard spheres, MRJ sphere packings}

\maketitle

\section{Introduction}
\label{sec:intro}

The statistics that structurally or topologically characterize the pore 
space of disordered porous media are intimately linked to their 
effective transport properties, such as the effective electrical 
conductivity $\sigma_e$~\cite{torquato_random_2002}, mean survival time 
$\overline{T}$~\cite{prager_interphase_1963-1,torquato_diffusion_1991, 
Note1}, principal (largest) diffusion relaxation time 
$T_1$~\cite{prager_interphase_1963-1,torquato_diffusion_1991}, and 
principal viscous relaxation time 
$\Theta_1$~\cite{avellaneda_rigorous_1991}.
For example, the first and second moments of the pore-size probability 
density function $P(\delta)$, $\langle \delta\rangle$ and $\langle 
\delta^2\rangle$, respectively, bound $\overline{T}$ and $T_1$ from above 
for diffusion-controlled reactions in which the interface of the porous 
medium is perfectly absorbing for a solute species diffusing in the pore 
space, where $P(\delta)d\delta$ is the probability that a randomly 
chosen point in the pore space lies at a distance between $\delta$ and 
$\delta+d\delta$ from the nearest point on the pore-solid 
interface~\cite{torquato_random_2002}.

An especially important pore characteristic is the \textit{critical pore 
radius} $\delta_c$ of a heterogeneous material, which is the maximal radius of 
an impenetrable test sphere so that the sphere can percolate through 
the void space.
Interestingly, as detailed below, the critical pore radius $\delta_c$ is 
related not only to all of the aforementioned effective transport 
properties ($\sigma_e,\overline{T}, T_1,\Theta_1$) of the porous medium
but also to its fluid permeability.

The \textit{fluid permeability} $k$ associated with slow viscous flow through an 
isotropic porous medium is defined by Darcy's law, which can be 
rigorously derived using homogenization 
theory~\cite{rubinstein_flow_1989}.
The permeability $k$ has dimensions of the square of length and, roughly 
speaking, may be regarded as an effective pore channel area of the 
``dynamically connected part of the pore 
space''~\cite{torquato_random_2002}.
Avellaneda and Torquato~\cite{avellaneda_rigorous_1991} used the 
solutions of unsteady Stokes equations for the fluid velocity vector 
field to derive a general rigorous relation connecting the fluid 
permeability $k$ to the formation factor $\cal F$ of the porous medium 
and a length scale $\cal L$ that is determined by the eigenvalues of the 
Stokes operator: 
\begin{equation}
 k=\frac{{\cal L}^2}{{\cal F}} \: , \label{k-exact}
\end{equation}
where $\cal L$ is a certain weighted sum over the {\it viscous 
relaxation times} $\Theta_n$ (i.e., inversely proportional to the 
eigenvalues of the Stokes operator), and ${\cal F}=\sigma_1/\sigma_e$ is 
the formation factor, where $\sigma_e$ is the effective electrical 
conductivity of a porous medium with a conducting fluid of conductivity 
$\sigma_1$ and a solid phase that is perfectly insulating.
Roughly speaking, the formation factor $\cal F$ quantifies the degree of 
``windiness'' for electrical transport pathways across a macroscopic 
sample~\cite{torquato_predicting_2020}.
(Note that the length scale $\cal L$ appearing in (\ref{k-exact}) 
absorbs a factor of 8 compared to the definition $L$ given in 
Ref.~\cite{avellaneda_rigorous_1991}; specifically, ${\cal L}=L/8$.) 

The prediction of the fluid permeability via theoretical methods is a 
notoriously difficult problem, largely because it is nontrivial to 
estimate the length scale ${\cal L}$ in (\ref{k-exact}) for general 
porous media.
Thus, the majority of previous analytical studies attempt to provide 
closed-form estimates of $\cal L$.
For example, the length scale $\cal L$ can be rigorously bounded from 
above by length scales associated with the mean survival time 
$\overline{T}$~\cite{torquato_relationship_1990}, principal diffusion relaxation 
time $T_1$~\cite{torquato_diffusion_1991}, and principal viscous 
relaxation time $\Theta_1$~\cite{avellaneda_rigorous_1991}.
There is a panoply of approximation formulas for $\cal 
L$~\cite{scheidegger_physics_1974, katz_quantitative_1986, 
johnson_new_1986, torquato_random_2002}.
An estimate due to Katz and Thompson~\cite{katz_quantitative_1986} 
approximates $\cal L$ to be proportional to the capillary radius at 
breakthrough during mercury injection in the pore space, which is 
directly related to the {\it critical pore radius} 
$\delta_c$~\cite{martys_length_1992}.
Empirical correlations between permeability and critical pore radius 
have also been found using the water expulsion 
method~\cite{nishiyama_permeability_2017}.

The critical pore radius is a complex structural characteristic that 
encodes both nontrivial geometrical and topological information.
Motivated by rigorous bounds on the principle relaxation time $T_1$ and 
its link to the permeability, \citet{torquato_predicting_2020} suggested 
the second moment of the pore size $\langle \delta^2 \rangle$ as an 
easily measurable approximation of ${\cal L}^2$ for models where the 
pore space is well connected.
The approximation was verified for BCC sphere packings~\cite{torquato_predicting_2020}.
Thus, $\langle \delta^2 \rangle$ is expected to be closely related to 
the critical pore radius, which we verify below.

Here, we study the critical pore radius and void percolation for 
disordered and ordered models of porous media derived from either 
overlapping or hard spheres (HS) with a constant radius $R$.
Such configurations of overlapping or hard spheres are effective models 
of a broad range of heterogeneous materials and many-particle 
systems~\cite{finney_modelling_1977, zallen_physics_1998, 
chaikin_principles_2000, manoharan_dense_2003, hansen_theory_2013, 
torquato_random_2002}.
Our models exhibit a varying degree of long- and short-range order, from 
completely random overlapping spheres to the crystalline densest packing 
of hard spheres.

Importantly, we determine the critical pore radius of maximally random 
jammed (MRJ) packings of identical spheres~\cite{torquato_is_2000}, 
which are, intuitively speaking, the maximally disordered among all 
mechanically stable packings.
More precisely, MRJ sphere packings minimize among jammed packings an 
order metric $\Psi$~\cite{torquato_is_2000, torquato_jammed_2010, 
ohern_random_2002, karayiannis_dense_2008, xu_maximally_2011, 
ozawa_jamming_2012, baranau_pore-size_2013, 
tian_geometric-structure_2015}.
Previously studied structural characteristics of MRJ sphere packings 
include their two-point statistics, average contact numbers, fractions 
of rattlers, Voronoi cell statistics and correlation functions, 
pore-size distributions, etc.~\cite{torquato_jammed_2010, 
jiao_nonuniversality_2011, atkinson_detailed_2013, 
klatt_characterization_2014, klatt_characterization_2016}.
Bounds on transport properties of MRJ packings have been recently 
characterized in Ref.~\cite{klatt_characterization_2018}.
\citet{ziff_percolation_2017} determined the site and bond percolation 
threshold of MRJ sphere packings.

We compare the critical pore radius of the MRJ sphere packings to three 
crystalline sphere packings and to three models with disordered 
microstructures.
The first model is that of overlapping spheres that are completely 
random and independent (also known as the Swiss-cheese 
model)~\cite{torquato_random_2002},
and the second model is that of equilibrium hard 
spheres~\cite{hansen_theory_2013}.
For the third model, we assign overlapping spheres to the points of 
amorphous inherent structures of the quantizer 
energy~\cite{klatt_universal_2019}, where the quantizer energy is 
proportional to the first moment of the void exclusion probability 
$E_V(r)$ (which is the probability that a randomly placed spherical 
cavity of radius $r$ contains no 
points)~\cite{torquato_reformulation_2010}.
Hence, the quantizer energy is also related to the pore-size 
distribution~\cite{klatt_characterization_2018}.
We therefore suggest it as an interesting model for studying transport 
properties.
For both the overlapping spheres (or Swiss-cheese model) and the 
quantizer model, we consider two different diameters of the spheres: (i) 
the average nearest-neighbor distance and (ii) diameters that result in 
the same porosity as MRJ sphere packings.

We quantify the degree of short-, intermediate-, and long-range order in 
our four systems using the $\tau$ order 
metric~\cite{torquato_ensemble_2015}.
It measures how the two-point statistics deviate from those of the 
Poisson point process: 
\begin{equation}
\begin{aligned}
  \tau &:=  \frac{1}{D^d}\int_{\mathbb{R}^d} [g_2(\vec{r})-1]^2d\vec{r}\\
       &\;= \frac{1}{(2\pi)^dD^d\rho^2}\int_{\mathbb{R}^d} 
  [S(\vec{k})-1]^2d\vec{k},
  \label{eq:tau_def}
\end{aligned}
\end{equation}
where $g_2(\vec{r})$ is the pair-correlation function and 
$S(\vec{k})$ the structure factor~\cite{hansen_theory_2013, 
torquato_random_2002}.
The systems are compared at unit number density
(with a cut-off value $k=16.5$ for the integration in Fourier space).

Here, we estimate the void percolation threshold using Kerstein's 
method~\cite{kerstein_equivalence_1983}, 
as described in Sec.~\ref{sec:methods}, and the Newman-Ziff 
algorithm~\cite{newman_fast_2001}.
The latter is based on the second moment of the cluster sizes and allows for a 
convenient finite-size scaling.

As mentioned above, Torquato \citet{torquato_predicting_2020} recently suggested suggested for porous 
media with a well-connected pore space to use the second moment of the 
pore size, $\langle\delta^2\rangle$, as a convenient estimate of 
$\mathcal{L}^2$, which in turn allows an estimation of the fluid 
permeability $k$.
Here, we compare the critical pore radius $\delta_c$ to 
$\langle\delta^2\rangle$ and confirm that to a good approximation 
$\delta_c^2\propto\langle\delta^2\rangle$.
In fact, we find that an estimation of $k$ based on $\langle \delta^2 
\rangle$ is superior to an estimate based on $\delta_c$ in that only the 
former provides the correct ranking of $k$ for our models.

We also compare the fluid permeability of models with different 
large-scale density fluctuations, i.e., nonhyperuniform and 
hyperuniform models.
A hyperuniform porous medium is defined by an anomalous suppression of 
long-wavelength volume-fraction fluctuations compared to those of 
typical disordered media~\cite{torquato_local_2003, 
torquato_hyperuniform_2018, zachary_hyperuniformity_2009}.
In agreement with the analysis of \citet{torquato_predicting_2020}, we 
find that the estimates of fluid permeabilities for our hyperuniform 
models tend to be smaller than those of their nonhyperuniform 
counterparts.

In the following, we first define our models, construction of Voronoi 
networks, and clustering analysis in Sec.~\ref{sec:methods}.
Then, we present our results on the critical pore radius, the pore 
statistics, and estimates of the fluid permeability in 
Sec.~\ref{sec:results}.
In Sec.~\ref{sec:conclusion}, we give concluding remarks and an outlook 
to future research.

\section{Models and structure characterization}
\label{sec:methods}

\begin{figure}[b]
  \centering 
  \includegraphics[width=\linewidth]{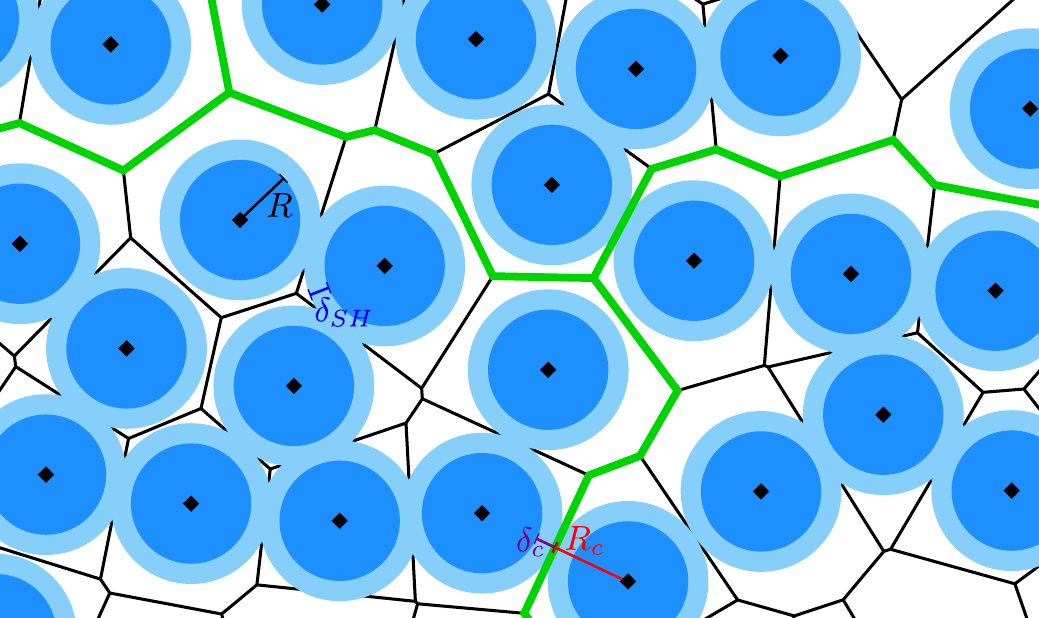}
  \caption{Two-dimensional schematic of how the critical pore size $\delta_c$ 
  of a dispersion of hard disks (dark blue or gray) can be determined from the 
  corresponding Voronoi diagram (black)~\cite{kerstein_equivalence_1983}.
  Each bond in the Voronoi diagram corresponds to a channel in the void 
  space.
  When each disk of radius $R$ is surrounded by a soft shell of 
  thickness $\delta_{SH}$ (light blue or gray),
  then $\delta_c$ is equal to the critical thickness at which the pore 
  space ceases to percolate;
  the thick (green) line highlights a cluster of open channels.}
  \label{fig:schematic}
\end{figure}

We use periodic boundary conditions for all of our samples, the 
construction of the Voronoi network, and the percolation analysis.
Figure~\ref{fig:schematic} schematically shows how the pore space is 
related to the Voronoi network.

\paragraph*{Models.}
The first model is that of overlapping spheres that are randomly and 
uniformly distributed in the simulation box without interaction.
Hence, the sphere centers are a snapshot of the ideal gas in the 
canonical ensemble, i.e., the number $N$ of points per sample is 
fixed.
Mathematically speaking, the points follow a binomial point process.
The $\tau$ order metric for this model is 0, by definition.

The second model that we study is an equilibrium fluid of hard spheres. 
The equal-sized spheres are impenetrable but do not interact 
otherwise.
Each sample has a packing fraction of 45\%.
The $\tau$ order metric is 9.45(1).

Determining the critical pore radius of equilibrium hard spheres is 
closely related to the so-called \textit{cherry-pit 
model}~\cite{torquato_random_2002}, where each hard sphere of radius $R$ 
is surrounded by a penetrable spherical shell of thickness $\delta_{SH}$.
The thickness at which the void space (outside the penetrable spheres) 
stops percolating is the critical pore radius $\delta_c$.
It is, therefore, strictly related to the void percolation threshold 
$R_c := R + \delta_c$.
The same principle applies to any other monodisperse sphere 
configuration; see Fig.~\ref{fig:schematic} for a two-dimensional 
schematic.

The third model is that of maximally random jammed (MRJ) packings of 
hard spheres.
Here, we analyze packings generated by \citet{atkinson_detailed_2013}.
The average packing fraction is 63.6\%.
The $\tau$ order metric is 23.7(1)~\cite{klatt_universal_2019}.

The fourth model is based on amorphous inherent structures of the 
quantizer energy~\cite{klatt_universal_2019}.
This energy functional is defined for Voronoi tessellations of arbitrary 
point configurations~\cite{liu_centroidal_2009, du_advances_2010, 
torquato_reformulation_2010, zhang_periodic_2012, ruscher_voronoi_2015, 
ruscher_voronoi_2018, klatt_universal_2019, ruscher_glassy_2020, 
hain_low-temperature_2020}, 
and it is proportional to a sum of the
second moments of inertia of all Voronoi cells (each computed
with respect to the corresponding Voronoi center).
The quantizer energy can be interpreted as a many-body interaction with 
a certain soft-core repulsion~\cite{torquato_reformulation_2010}.
It has been studied both as a ground-state 
problem~\cite{torquato_reformulation_2010}
and at finite 
temperature~\cite{ruscher_voronoi_2015,ruscher_voronoi_2018,ruscher_glassy_2020, 
hain_low-temperature_2020}.

More precisely, the quantizer energy can be defined as the first moment 
of the void exclusion probability $E_V(r)$ of the point configuration.
For a point pattern at unit number density, the rescaled quantizer 
energy (or error) is given by~\cite{torquato_reformulation_2010}:
\begin{align}
  \mathcal{G} := \frac{1}{d}\langle r^2\rangle = \frac{2}{d}\int_0^{\infty}rE_V(r)dr,
  \label{eq:def_quant}
\end{align}
where $d$ is the dimension (here $d=3$).
For monodisperse sphere packings with radius $R$, the 
complementary cumulative distribution function $F(\delta)$ of the pore 
size is trivially related to the exclusion probability via 
$E_V(r)=\phi_1F(r-R)$ for $r > R$, where $\phi_1$ is the volume 
fraction of the pore space (and $\phi_2=1-\phi_1$ is the volume fraction 
of the spheres)~\cite{torquato_random_2002}.
Hence, the quantizer energy is closely related to the second moment of 
the pore size $\langle \delta^2 \rangle = 2 \int_0^{\infty}\delta 
F(\delta)d\delta$;
in fact, for point particles with $R=0$: $\mathcal{G}=\langle \delta^2 
\rangle/d$; and for nonoverlapping spheres, the following relation can 
be straightforwardly derived by using Eq.~(5.68) in 
\citet{torquato_random_2002}:
\begin{align}
  \mathcal{G}=\frac{3+2\phi_1}{5d}R^2+\frac{2\phi_1}{d}R\langle\delta\rangle+\frac{\phi_1}{d}\langle\delta^2 
  \rangle.
  \label{eq:quantizer_and_delta}
\end{align}
Optimizing the quantizer energy for the centers of a sphere packing 
is, therefore, closely related to an optimization of its pore 
statistics.

\begin{figure*}[t]
  \centering
  \includegraphics[width=0.38\textwidth,valign=t]{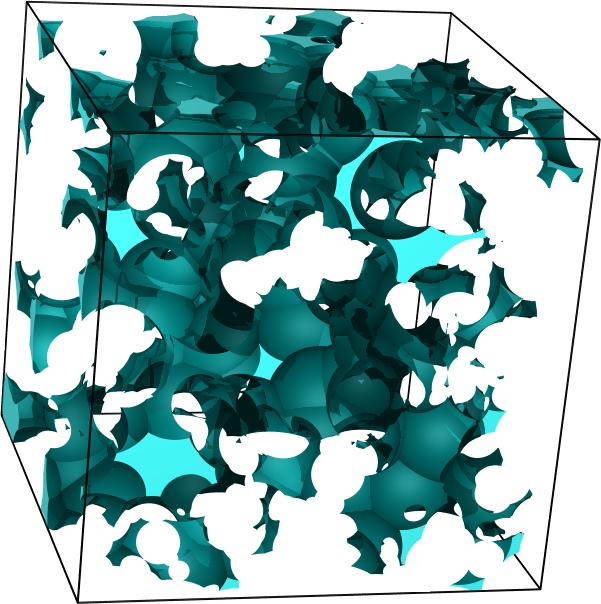}\hfill%
  \includegraphics[width=0.58\textwidth,valign=t]{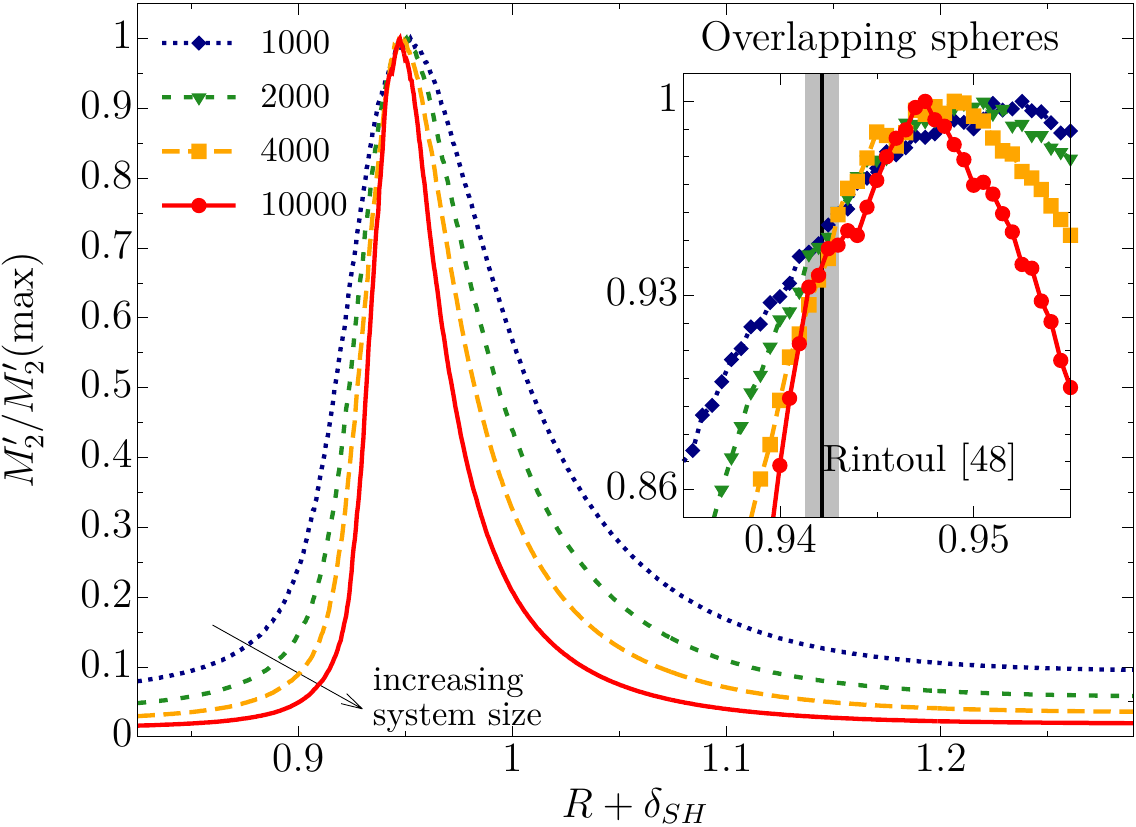}
  \caption{Overlapping spheres: a sample of the void space  (left) 
  and the rescaled cluster index $M_2’$ as a function of the sphere 
  radii $R+\delta_{SH}$.
  The curves for different system sizes intersect (see inset) roughly in 
  one point, which corresponds to the percolation threshold $R_c$.
  The value agrees within statistical accuracy with the previous result 
  by~\citet{rintoul_precise_2000}, where the mean value is indicated by 
  the vertical line and the error by the gray band.
  In the sample of the void space the different colors (shades) are only 
  for an improved three-dimensional visualization.}
  \label{fig:ideal_gas}
\end{figure*}

\begin{figure*}[t]
  \centering
  \includegraphics[width=0.38\textwidth,valign=t]{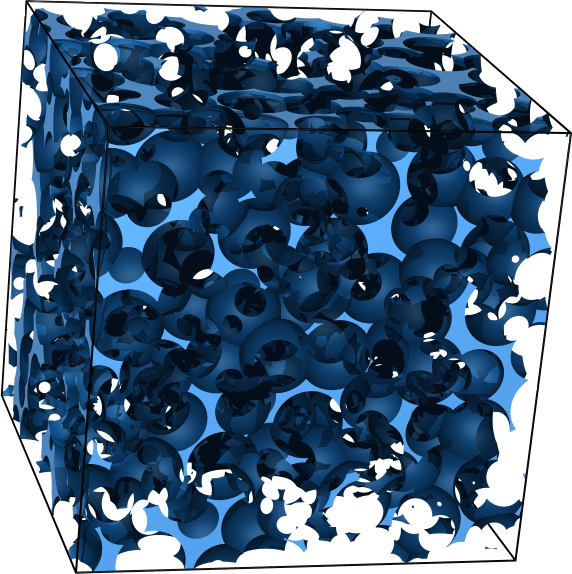}\hfill%
  \includegraphics[width=0.58\textwidth,valign=t]{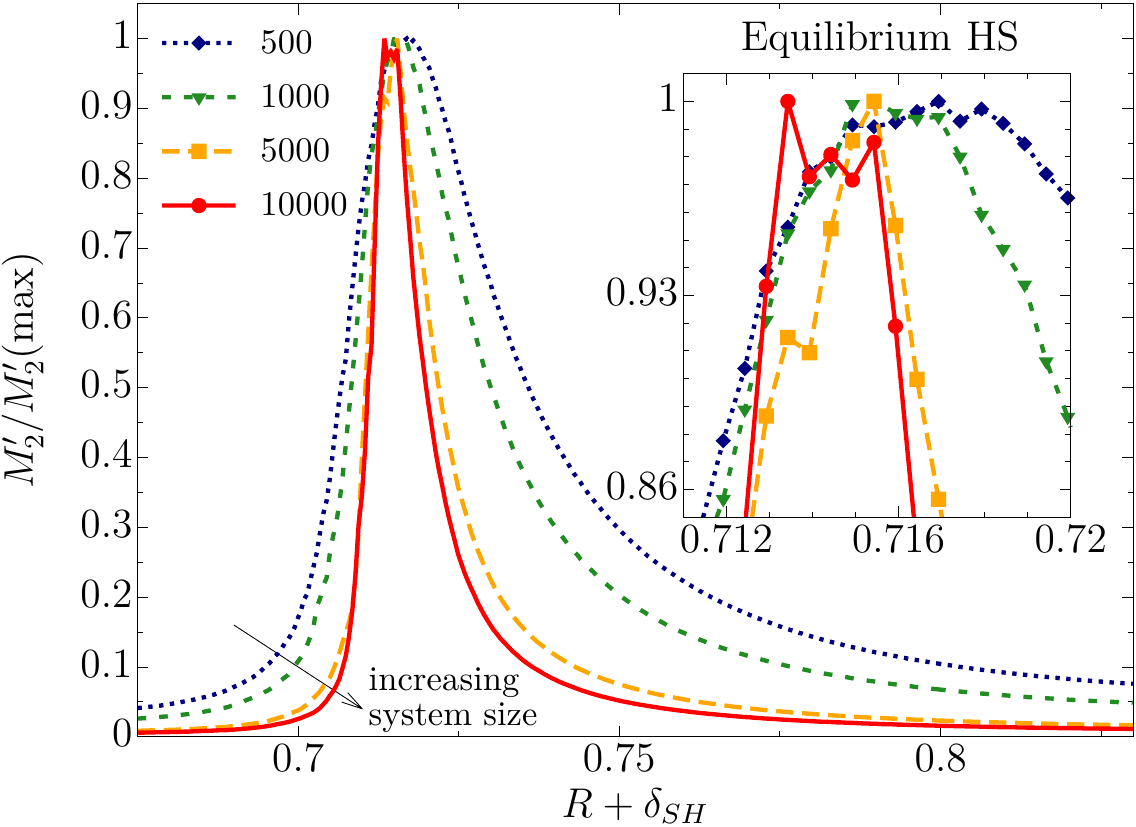}
  \caption{Equilibrium hard spheres: a sample of the void space
  surrounding the dilated, ``cherry-pit'' spheres with radius $R+\delta_{SH}$ (left) 
  and the rescaled cluster index $M_2’$ as a function of the sphere 
  radii $R+\delta_{SH}$; for more details, see Fig.~\ref{fig:ideal_gas}.}
  \label{fig:equilibrium}
\end{figure*}

\begin{figure*}[t]
  \centering
  \includegraphics[width=0.38\textwidth,valign=t]{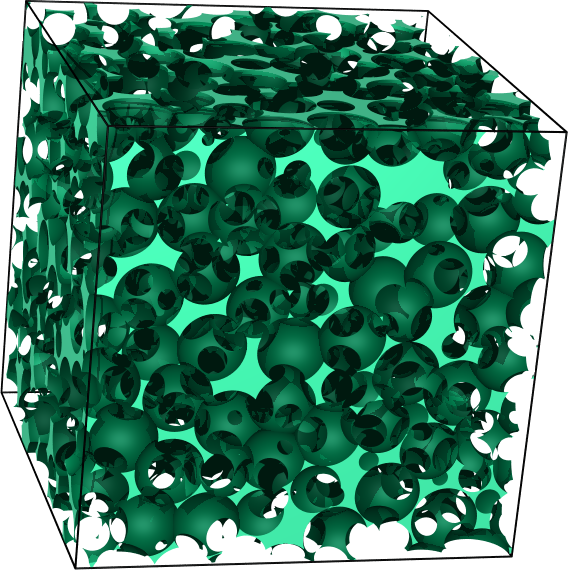}\hfill%
  \includegraphics[width=0.58\textwidth,valign=t]{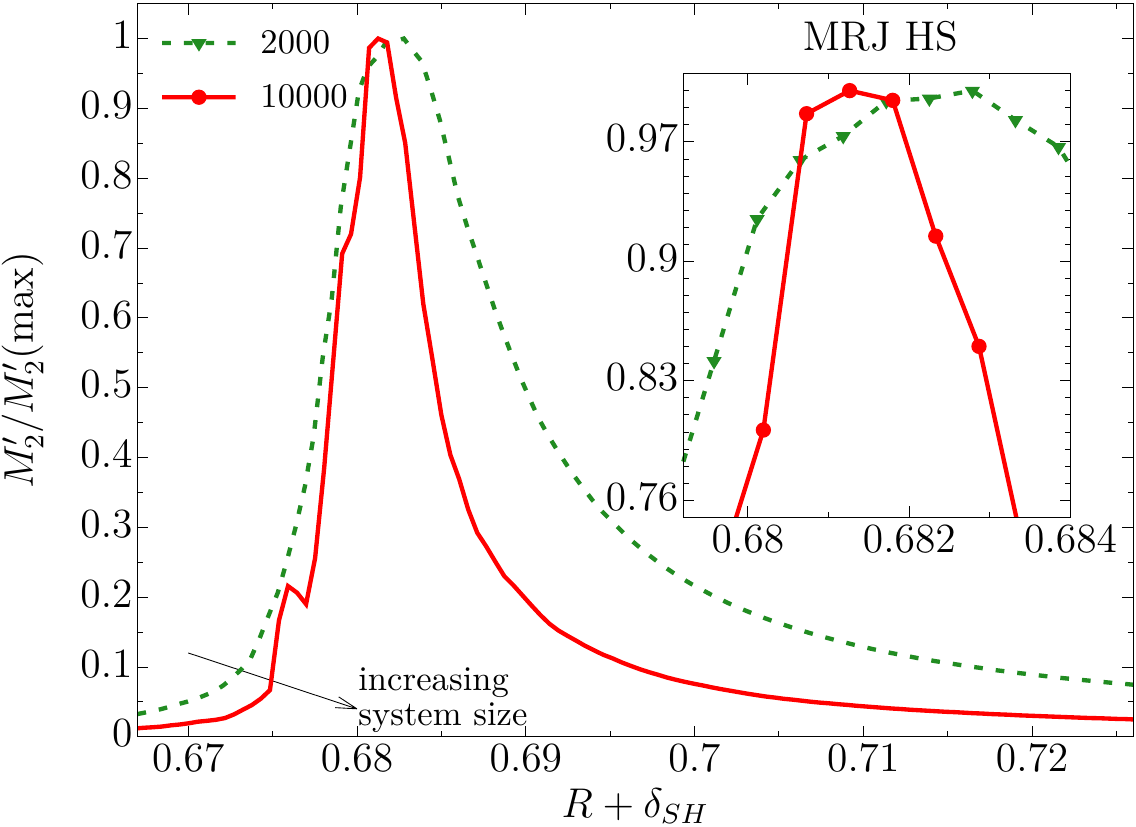}
  \caption{MRJ hard spheres: a sample of the void space
  surrounding the dilated, ``cherry-pit'' spheres with radius $R+\delta_{SH}$ (left) 
  and the rescaled cluster index $M_2’$ as a function of the sphere 
  radii $R+\delta_{SH}$; for more details, see Fig.~\ref{fig:ideal_gas}.}
  \label{fig:mrj}
\end{figure*}

\begin{figure*}[t]
  \centering
  \includegraphics[width=0.38\textwidth,valign=t]{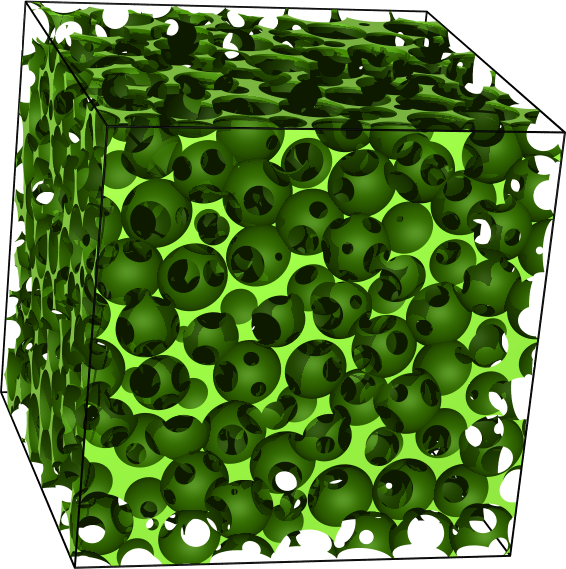}\hfill%
  \includegraphics[width=0.58\textwidth,valign=t]{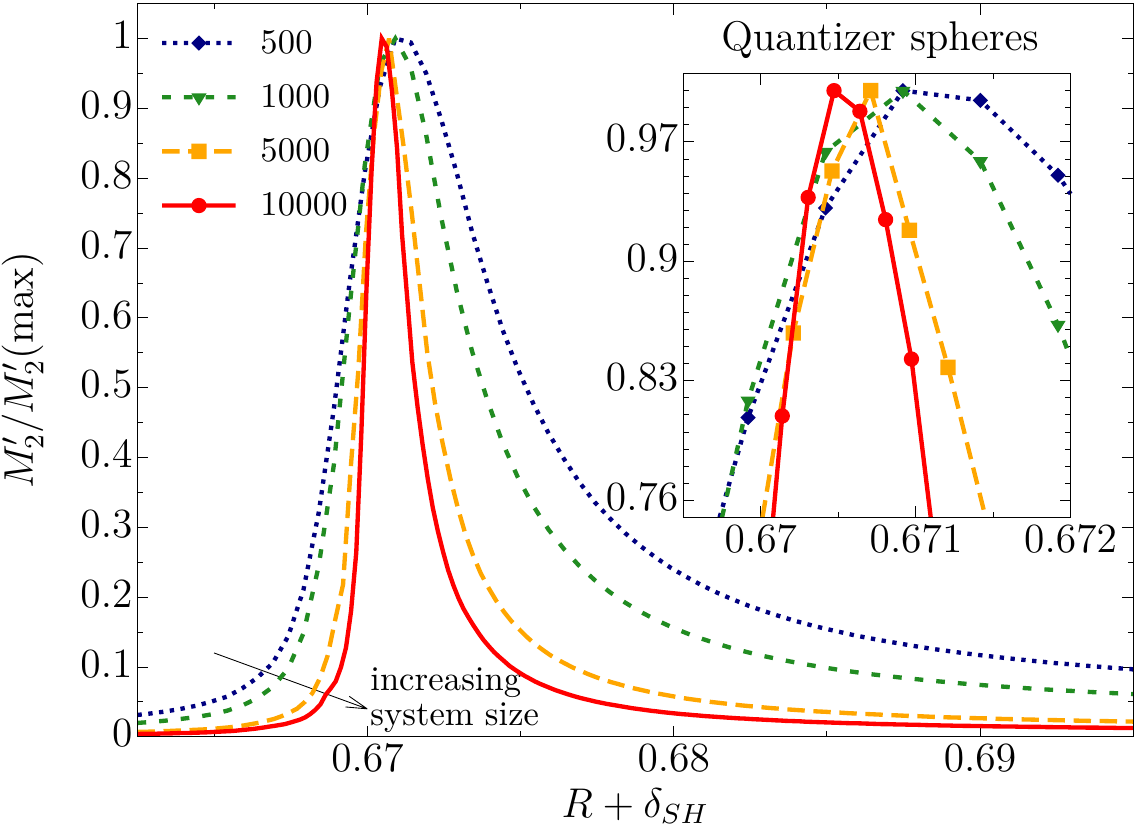}
  \caption{Quantizer spheres: a sample of the void space
  surrounding the dilated, ``cherry-pit'' spheres with radius $R+\delta_{SH}$ (left) 
  and the rescaled cluster index $M_2’$ as a function of the sphere 
  radii $R+\delta_{SH}$; for more details, see Fig.~\ref{fig:ideal_gas}.}
  \label{fig:quantizer}
\end{figure*}

\begin{figure*}[t]
  \centering
  \includegraphics[width=\textwidth]{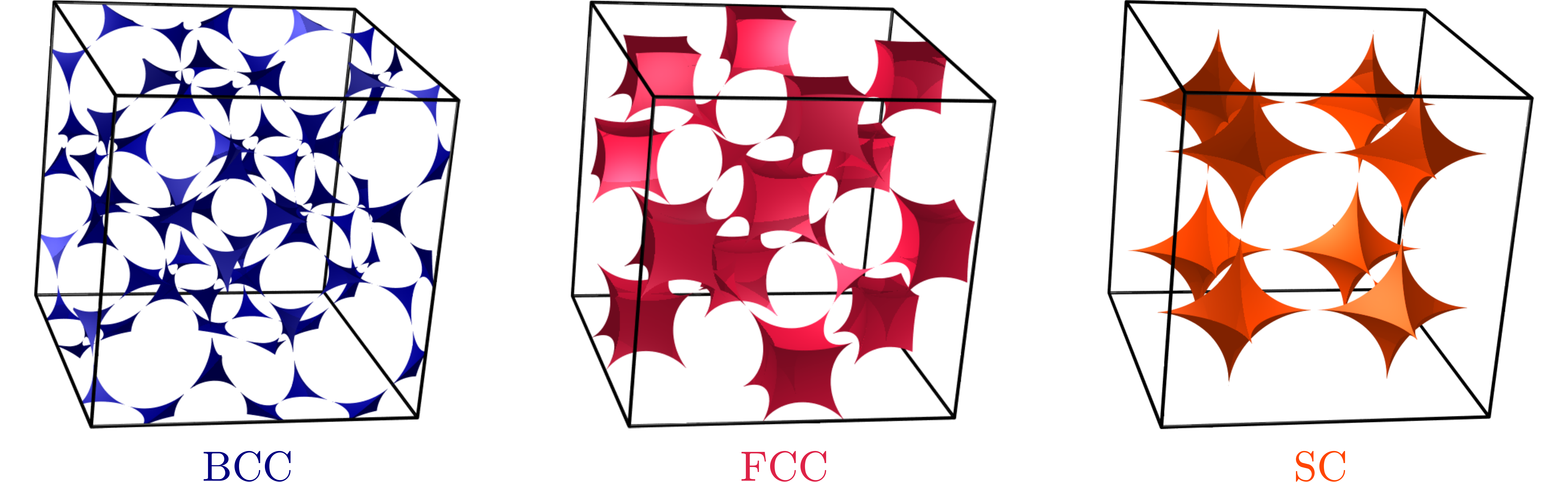}
  \caption{Void space surrounding the soft overlapping shells at the 
  critical point for BCC (left), FCC (center), and SC (right) 
  crystalline sphere configurations:
  The different colors (shades) are only for an improved 
  three-dimensional visualization. The critical porosity 
  $\varphi_c:=E_V(R_c)$ is distinctly smaller for the BCC spheres than for the FCC 
  and SC spheres.}
  \label{fig:crystals}
\end{figure*}

To construct our samples of amorphous inherent structures, we start from 
a binomial point process and locally minimize the quantizer energy using 
the Lloyd algorithm~\cite{klatt_universal_2019}.
In each step of the algorithm and for each cell, the Voronoi center is 
replaced by the center of mass of the cell~\cite{lloyd_least_1982}.
We apply 10,000 steps, after which the algorithm converges to an 
amorphous inherent structure with a strong suppression of density 
fluctuations~\cite{klatt_universal_2019}.
The final states are (effectively) centroidal Voronoi tessellations, 
where in each cell the Voronoi center coincides with the center of mass.
The quantizer energy of the disordered inherent structures 
($\mathcal{G}=0.07917$) is only slightly larger than that of the 
(conjectured) crystalline ground-state, the body-centered cubic (BCC) 
lattice ($\mathcal{G}=0.07854$)~\cite{klatt_universal_2019}.
The $\tau$ order metric of the amorphous inherent structures of the 
quantizer energy is 31.6(2)~\cite{klatt_universal_2019}, i.e., larger 
than the value for MRJ sphere packings by about a factor of $4/3$.

The corresponding ground-state problem, known as the ``quantizer 
problem,'' is also related to another tessellation optimization problem, 
known as the ``covering problem''~\cite{torquato_reformulation_2010}.
The latter problem is the search for a point configuration that 
minimizes the radius of overlapping circumscribed spheres to cover the 
space.
This covering radius is always an upper bound on the critical radius 
$R_c$.
Since MRJ sphere packings are saturated, they have a finite covering 
radius, like a crystal.
A finite covering radius $R_{\text{cov}}<\infty$ implies that the 
exclusion probability $E_V(r)$ has compact support, specifically, 
$E_V(r)=0$ for $r\geq R_{\text{cov}}$.
As for the quantizer problem, the BCC lattice is believed to be the 
optimum of the covering problem.
Both the quantizer and covering problems have relevance in numerous 
applications, from wireless communication and network layouts, to data 
compression and cryptography; see \citet{torquato_reformulation_2010} 
and references therein.

We compare the pore statistics and transport properties of our 
four disordered models to those of three perfectly ordered crystalline 
structures.
Specifically, we here consider dense lattice packings of spheres with simple cubic 
(SC), body-centered cubic (BCC), and face-centered cubic (FCC) 
symmetries.

\paragraph*{Simulation details.}
The MRJ samples are simulated in unit cells with a nonorthogonal basis.
All other samples are simulated in cubic unit cells.
Table~\ref{tab:num_samples} lists the number of samples and the number of 
points per sample.
The total number of points in our samples is more than $80\times10^6$.
The number density $\rho$ is the average number of points per unit 
volume.
We choose the unit of the length such that $\rho=1$ for all of our 
models.
For the overlapping and equilibrium hard spheres and for the quantizer sphere configurations, the 
number density is fixed for each sample.
For the MRJ spheres packings, the radius of the spheres is fixed, but 
the number density slightly fluctuates around unity.
Figures~\ref{fig:ideal_gas}--\ref{fig:crystals} show for each of our models a sample of the void 
space.

\begin{table}[b]
  \centering
  \begin{tabular}{
    l r c r l l
    l r c r l l}
    \toprule
    \multicolumn{6}{c}{Overlapping spheres}&
    \multicolumn{6}{c}{Equilibrium HS}\\
    \cline{2-5}
    \cline{8-11}
        & $20,000$ & $\times$ & $1000  $ & points&
      & & $2000  $ & $\times$ & $500   $ & points&\\
        & $10,000$ & $\times$ & $2000  $ & points&
      & & $1000  $ & $\times$ & $1000  $ & points&\\
        & $5000  $ & $\times$ & $4000  $ & points&
      & & $200   $ & $\times$ & $5000  $ & points&\\
        & $1000  $ & $\times$ & $10,000$ & points&
      & & $100   $ & $\times$ & $10,000$ & points&\\
      \\
    \multicolumn{6}{c}{MRJ HS}&
    \multicolumn{6}{c}{Quantizer spheres}\\
    \cline{2-5}
    \cline{8-11}
        &          &          &          &       &
      & & $2000  $ & $\times$ & $500   $ & points&\\
        & $1015  $ & $\times$ & $2000  $ & points&
      & & $1000  $ & $\times$ & $1000  $ & points&\\
        &          &          &          &       &
      & & $200   $ & $\times$ & $5000  $ & points&\\
        & $16    $ & $\times$ & $10,000$ & points&
      & & $100   $ & $\times$ & $10,000$ & points&\\
    \toprule
  \end{tabular}
  \caption{The number of samples and number points per sample for each 
  of our models.}
  \label{tab:num_samples}
\end{table}

\paragraph*{Voronoi network.}
For monodisperse sphere configurations, the void percolation can be 
accurately studied by reformulating it as a ``weighted bond 
percolation'' on the Voronoi network, as discussed by 
Kerstein~\cite{kerstein_equivalence_1983}; see Fig.~\ref{fig:schematic}.
The topology of the void space is related to that of the Voronoi 
network, i.e., the network formed by the edges of the Voronoi diagram.
Each channel in the void space corresponds to a bond in the Voronoi 
network.
The channel vanishes when $R+\delta_{SH}$ is larger or equal to the 
distance of the bond to its Voronoi neighbors.
Kerstein's method~\cite{kerstein_equivalence_1983} has been previously 
used to study void percolation for overlapping 
spheres~\cite{elam_critical_1984, rintoul_precise_2000, 
hofling_critical_2008, priour_percolation_2018}
and hard-sphere packings (both jammed and in 
equilibrium)~\cite{spanner_splitting_2016},
including models for protein structures~\cite{treado_void_2019}.

Following this idea by \citet{kerstein_equivalence_1983}, we construct 
for each sample the Voronoi diagrams using 
\textsc{voro++}~\cite{rycroft_analysis_2006,rycroft_voro_2009}.
By identifying vertices within an accuracy of about $10^{-12}$, we 
determine the Voronoi network (of cell edges) and assign to each edge 
the smallest distance to its neighboring Voronoi centers.
The void percolation problem is thus equivalent 
to a weighted bond percolation problem on the Voronoi network.

\paragraph*{Newman-Ziff method.}

Our goal was to find the critical percolation threshold of the void 
system.
Along each bond in the Voronoi network, we assigned a weight equal to the 
distance of the bond to the neighboring Voronoi centers (which is 
directly related to the radius of a sphere that can just pass through 
that pore throat), and the goal is to find the critical radius of a 
sphere where the system percolates.
Various criteria can be used to determine the percolation point.
A common one has been the point where a single cluster of connected 
vertices spans from one side of the system to the other, or for a periodic system, 
where it wraps around.
However, various other criteria can be used, including Binder-type 
ratios~\cite{wang_bond_2013} involving moments 
of the size of the largest cluster.
The goal in these is to find something universal so that its 
value is independent of the size of the system under finite-size scaling 
(although the corrections to scaling will cause a size dependence 
visible for smaller systems.)   Here we use another universal quantity: 
the second moment of the size distribution leaving out the largest 
cluster, $M_2’$, divided by its value at the maximum of the curve 
$M_2’(\max)$.
The idea behind this is that $M_2’$ is a peaked function whose peak is 
near the percolation threshold $p_c$ but not exactly at $p_c$.
Finite-size scaling theory implies that $L^{-\gamma/\nu} M_2’(p)$ 
becomes a function of $(p-p_c)L^{1/\nu}$ in the scaling limits $p \to 
p_c$ and $L \to \infty$.
If we divide the value at $p_c$ by the value at the maximum, for 
example, then we get a ratio, which is universal (the same for all 
systems of the same dimensionality and shape). Thus, if we consider 
plots of $M_2’(p)/M_2’(\max)$, the crossing of the curves will 
indicate the critical point.
For very precise determinations of the critical point, one would also 
have to worry about the corrections-to-scaling contribution, but to the 
precision available for the systems here, this is not necessary.

In the Newman-Ziff algorithm, bonds are added one at a time, and the 
union-find computer science algorithm is used to keep track of the 
evolving cluster size distribution in a very efficient manner, including 
the moments such as $M_2’$.
This is, in fact, much easier than determining crossing or wrapping, which 
requires extra components in the data structure.
Before carrying out the algorithm, we sort all the bonds from large to 
small weights and then add the bonds one at a time (largest weights 
first).
Thus, for a given sample of a lattice, we could only carry out one test 
of the percolation threshold, unlike in typical lattices where we could 
create many measurements by occupying the bonds in random order.
Here the order of the bonds is fixed by their weight.
The algorithm works in a ``microcanonical'' space where averaged 
quantities are determined as a function of the number of bonds made 
occupied.
Usually, one carries out a convolution of the microcanonical 
measurements with a binomial distribution to get the ``canonical'' 
behavior that gives results as a function of $p$.
Here we do not do that, because for one thing there is no random bond 
occupation probability $p$ here, and, secondly, the difference between the 
two is slight and would not be observable with the precision of the 
results that we are able to get here.
Since $M_2’$ can strongly fluctuate between samples, we first 
bin, for each model and system size separately, all weights that we find 
(using a constant bin width).
Then, we average the corresponding values of $M_2’$ within each bin.

The crossing point of $M_2’/M_2’(\max)$ is used to find the threshold.
Its value should be universal and the same for all systems of the same 
shape and boundary conditions.
We verified this by considering bond percolation on the simple cubic 
lattice, and confirm the threshold of $p_c = 0.24881$ with a crossing 
point of $M_2’/M_2’(\max) = 0.96$, consistent with the values found here 
(about 0.94--0.96) for these quite different systems.

\paragraph*{Pore size.}
The pore size $\delta$ can be easily estimated from both simulated data 
and three-dimensional images of real porous 
media~\cite{coker_morphology_1996}.
Here, we determine the mean pore size $\langle \delta \rangle$ and the 
second moment of the pore size $\langle \delta^2 \rangle$ using a 
straightforward Monte Carlo sampling.
Points are placed randomly and uniformly distributed in the pore space 
surrounding the spheres.
For each point, we determine the smallest distance to a sphere and 
estimate the first and second moment of $\delta$ using the arithmetic 
mean.
We estimate the statistical error using the standard error of the mean.
The number of Monte Carlo points per sample is $10^5$, where for each 
model we analyze each sample of the two largest system sizes.
For overlapping spheres, the pore-size distribution is known 
analytically.
We also determine the pore sizes for lattice packings of spheres, where 
we use $10^7$ sampling points for each lattice.
For the tabulated values of $\langle\delta\rangle$ and 
$\langle\delta^2\rangle$ of the dense hard-sphere lattice packings, we 
use the values from Eqs.~(B24)--(B39) in 
Ref.~\cite{klatt_characterization_2018}, which were obtained by 
numerical integration of exact formulas~\cite{Note3}.
For the SC and BCC sphere packings, we also confirm these values by 
numerical integration of the exact formulas for $E_V(R)$ from Eqs.~(84) 
and (87) in Ref.~\cite{torquato_reformulation_2010}.
An exact formula of $E_V(R)$ also allows for precise values of the 
critical void porosity $\varphi_c := E_V(R_c)$ of the void space 
surrounding the soft shells at the critical radius.

\begin{table*}[t]
  \centering
  \addtolength{\tabcolsep}{8pt}
  \begin{tabular}{l l c l c l@{\,}l}
   \toprule
    Model                           &
    \multicolumn{1}{c}{$\phi_1$}    &
    \multicolumn{1}{c}{$R_c$}       &
    \multicolumn{1}{c}{$\varphi_c$} &
    \multicolumn{1}{c}{$\delta_c$}  &
    \multicolumn{2}{c}{$\langle\delta^2\rangle$} \\
    \hline                                                   
    Overlapping spheres & 0.9148\dots & 0.943(3)   & 0.0298(10)  & 0.666(3)   & 1.274\dots           & $\times 10^{-1}$ \\
                        & 0.3640\dots & 0.943(3)   & 0.0298(10)  & 0.320(3)   & 3.321\dots           & $\times 10^{-2}$ \\
    Equilibrium HS      & 0.550       & 0.714(2)   & 0.0257(12)  & 0.239(2)   & 1.5562(5)            & $\times 10^{-2}$ \\
    SC HS               & 0.4764\dots & 0.707\dots & 0.0349\dots & 0.207\dots & 1.388\dots      & $\times 10^{-2}$ \\
    Quantizer spheres   & 0.430       & 0.670(1)   & 0.0179(8)   & 0.156(1)   & 6.793(2)\hphantom{1} & $\times 10^{-3}$ \\
                        & 0.364       & 0.670(1)   & 0.0179(8)   & 0.136(1)   & 5.269(2)\hphantom{1} & $\times 10^{-3}$ \\
    MRJ HS              & 0.364       & 0.681(2)   & 0.0303(16)  & 0.148(2)   & 7.177(2)\hphantom{1} & $\times 10^{-3}$ \\
    BCC HS              & 0.3198\dots & 0.668\dots & 0.0055\dots & 0.122\dots & 3.718\dots      & $\times 10^{-3}$ \\
    FCC HS              & 0.2595\dots & 0.648\dots & 0.0358(6)   & 0.086\dots & 3.592\dots      & $\times 10^{-3}$ \\
    \botrule
  \end{tabular}
  \caption{
  Void percolation and pore-size statistics for our models with 
  different porosities $\phi_1$: The table shows the critical radius 
  $R_c$, the critical void porosity $\varphi_c$ (surrounding the soft 
  shells at the critical point), and it compares the critical pore 
  radius $\delta_c$ to $\langle\delta^2\rangle$.
  While $R_c$ and $\varphi_c$ only depend on the positions of the sphere 
  centers, $\delta_c$ and $\langle\delta^2\rangle$ also depend on the 
  radius $R$ and hence on the porosity $\phi_1$.
  For the overlapping spheres and the quantizer model, the table shows 
  the values for two different porosities: for a radius that matches 
  half the nearest-neighbor distance and for a porosity that matches the 
  MRJ value.}
  \label{tab:critical}
\end{table*}

\section{Results}
\label{sec:results}

Figures~\ref{fig:ideal_gas}--\ref{fig:quantizer} show for our four 
models of disordered sphere configurations the curves of the rescaled 
cluster index $M_2’$ as a function of the sphere radii $R+\delta_{SH}$.
The insets zoom into the region where the curves of different system 
sizes intersect, i.e., at the percolation threshold $R_c$ in the 
infinite-system size limit (where $R_c$ is the critical value of 
$R+\delta_{SH}$).
Each figure also shows a sample of the void space for radii below the 
percolation threshold (about 90\% of $R_c$).

Table~\ref{tab:critical} lists our estimates of the critical radius 
$R_c$, critical void porosity $\varphi_c := E_V(R_c)$, and critical pore 
radius $\delta_c$.
The table compares the values for the disordered sphere 
configurations to those of the crystalline sphere configurations.
Our results for $R_c$ and $\varphi_c$ of the overlapping sphere model 
and equilibrium hard spheres agree within statistical errors with 
previous results~\cite{elam_critical_1984, rintoul_precise_2000, 
hofling_critical_2008, spanner_splitting_2016, priour_percolation_2018, 
Note2}.
In particular, for the overlapping spheres
$R_c = 0.943(3)$ and $\varphi_c = 0.0298(10)$
agree with the estimates 
$R_c = 0.942(1)$ and $\varphi_c = 0.0301(3)$
by \citet{rintoul_precise_2000},
$R_c = 0.9425(9)$ and $\varphi_c = 0.0300(3)$
by \citet{hofling_critical_2008},
and
$R_c = 0.9422(3)$ and $\varphi_c = 0.0301(1)$
by \citet{priour_percolation_2018}.
Moreover, our estimate $R_c = 0.714(2)$ for  the equilibrium hard 
spheres agrees with the estimate of $R_c = 0.712(4)$ that we obtain from 
Fig.~S1 in \citet{spanner_splitting_2016}.

\begin{figure}[b]
  \centering 
  \includegraphics[width=\linewidth]{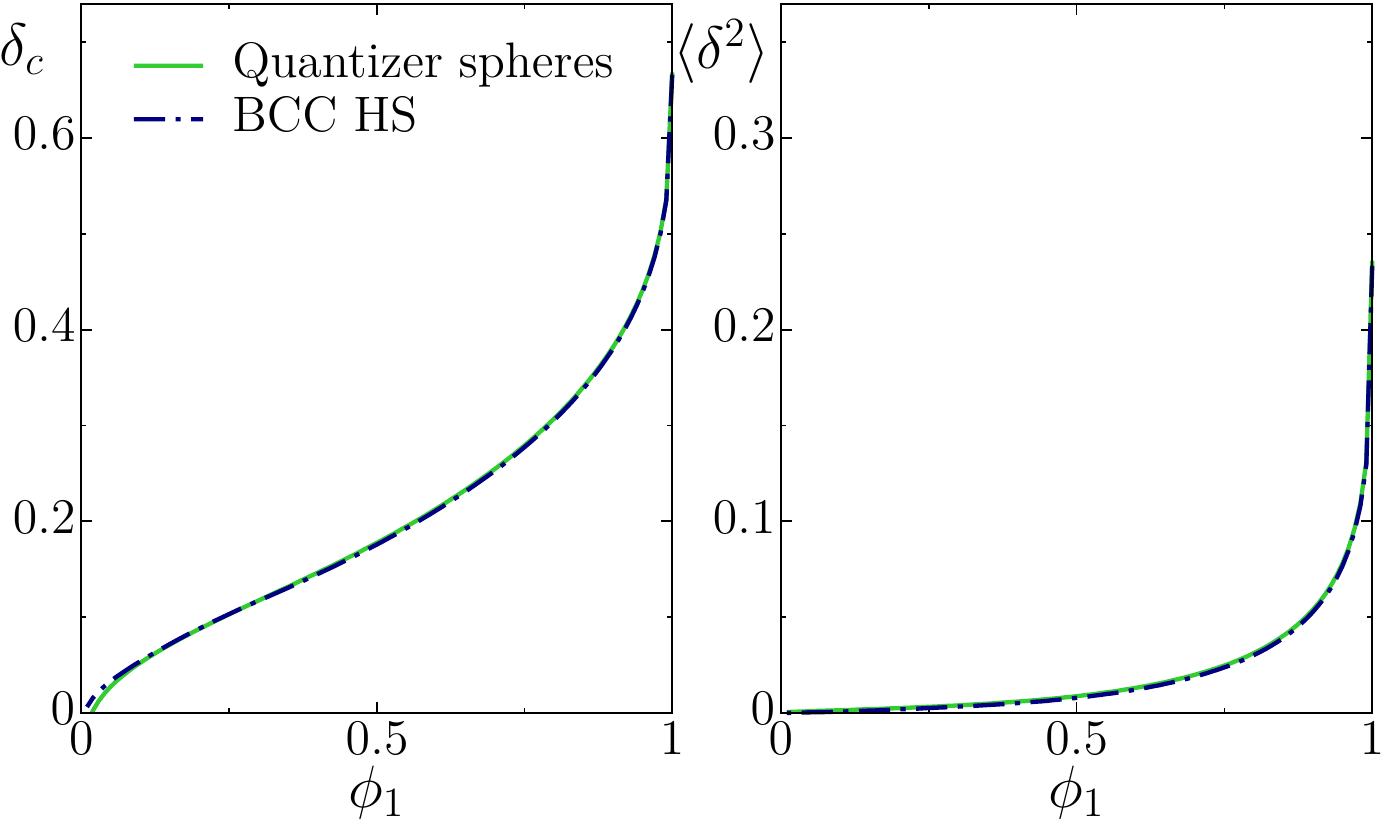}
  \caption{The critical pore radius, $\delta_c$ (left), and the second moment of 
  the pore size, $\langle\delta^2\rangle$ (right), are compared for dispersions of 
  spheres arranged either on a BCC lattice or according to our amorphous 
  inherent structures of the quantizer energy.}
  \label{fig:bccvsquantizer}
\end{figure}

For our disordered sphere models, we find that the percolation threshold $R_c$ decreases 
with increasing order, as measured by the $\tau$ order metric.
Moreover, while the amorphous hard-sphere packings have a distinctly 
larger value of $R_c$ than the optimal FCC packing, the amorphous 
quantizer states have about the same $R_c$ as the (conjectured) optimal 
quantizer, a BCC lattice.
The values agree within 0.3\%.
For the corresponding dispersions of spheres, we find for all radii considered here that the 
second moment of the pore size, $\langle\delta^2\rangle$ agrees within 
0.2\% (even if the spheres overlap); see Fig.~\ref{fig:bccvsquantizer}.

Among the disordered models, the critical porosity $\varphi_c$ is lowest 
for the quantizer spheres [$0.0179(8)$].
For the lattices, the lowest value is attained by the BCC lattice 
($0.0055\dots$).
In contrast, the critical porosity of the FCC lattice [$0.0358(6)$] 
is even larger than that of overlapping spheres [$0.0298(10)$].
The large difference between $\varphi_c$ for BCC and FCC lattices is 
related to the shape of the holes between the overlapping soft sphere 
shells.
For the BCC lattice, there are only small, so-called `tetrahedral' 
holes, but for the FCC lattice, there is an additional, relatively large 
type of hole,  called `octahedral.'
These octahedral holes are formed by six neighboring spheres, whose 
centers form a regular octahedron; the interstice between the spheres 
has a shape that resembles a cube (which is the dual polyhedron of an 
octahedron); see Fig.~\ref{fig:crystals}.

Next, we compare the critical pore radius to the pore-size statistics. 
We have found the following mean pore sizes (compared at 
unit number density):
for overlapping spheres, $\langle \delta\rangle=
0.30933\dots$ at $\phi_1=0.9148\dots$
and $\langle \delta\rangle=
0.14346\dots$ at $\phi_1=0.3640\dots$;
for equilibrium hard spheres, $\langle \delta\rangle=
0.10259(2)$;
for SC HS, $\langle \delta\rangle=
0.09602\dots$;
for quantizer spheres, $\langle \delta\rangle=
0.06881(1)$ at $\phi_1=0.430$
and $\langle \delta\rangle=
0.05990(1)$ at $\phi_1=0.364$;
for MRJ HS, $\langle \delta\rangle=
0.067414(8)$;
for BCC HS, $\langle \delta\rangle=
0.05095\dots$; and
for FCC HS, $\langle \delta\rangle=
0.04674\dots$.
Table~\ref{tab:critical} lists the second moments of the pore sizes for 
our models.

\begin{figure}[b]
  \centering 
  \includegraphics[width=\linewidth]{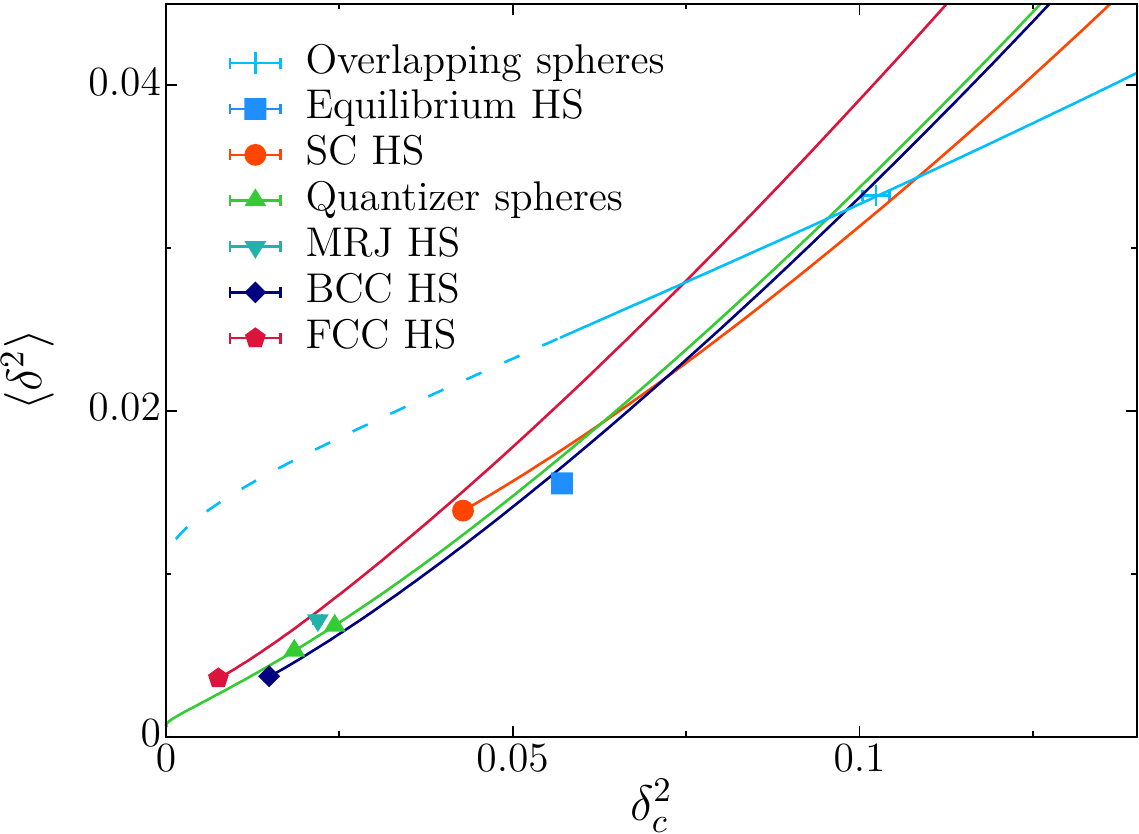}
  \caption{Comparison of the square of the critical pore radius 
  $\delta_c^2$ and the second moment of the pore-size distribution 
  $\delta^2$:
  The curves represent different values of $\delta_{SH}$ with fixed 
  sphere centers.
  For precise values, see Table~\ref{tab:critical}.
  The solid and dashed lines represent overlapping spheres with 
  porosities above and below 0.23.
  For the entire range of our models and porosities, we find 
  $\delta_c^2\propto\delta^2$ to a good approximation for models with a 
  well-connected pore space.
  However, as shown in Figs.~\ref{fig:permeability} and 
  \ref{fig:permeability-delta_sq}, $\delta_c^2$ and $\delta^2$ predict 
  different rankings of the fluid permeability $k$ for dispersions 
  of spheres at a given volume fraction.}
  \label{fig:cf-delta}
\end{figure}

Following the suggestion by \citet{torquato_predicting_2020}, 
Fig.~\ref{fig:cf-delta} compares the square of the critical pore radius 
to the second moments of the pore size.
To compare the models for a broad range of porosities,
we here vary the sphere radii $R$ for each model (from $0$ to 
$\infty$).
In agreement with the suggestion, we find that $\delta_c$ is, to a good 
approximation,
proportional to $\langle \delta^2 \rangle$ over our entire range of 
models and porosities.
However, $\delta_c^2$ and $\delta^2$ lead to different predictions of 
the rankings of the fluid permeability $k$ for dispersions of 
spheres at a given volume fraction, as discussed below.

The approximation of $\mathcal{L}^2$ by $\langle \delta^2 \rangle$ was 
suggested by \citet{torquato_predicting_2020} for models in which the 
pore space is well connected.
We, therefore, distinguish between overlapping sphere configurations 
above (solid line) and below (dashed line) a porosity of 0.23.
The approximation is most accurate for overlapping spheres if the 
porosity is similar to that of MRJ spheres.

For an estimate of the fluid permeability $k$, we additionally need to 
approximate the formation factor $\mathcal{F}$.
\citet{torquato_effective_1985} derived a tight lower bound on 
$\mathcal{F}$ for any three-dimensional porous medium that accounts for up to 
four-point information.
For both ordered and disordered dispersions of particles, the four-point 
parameter vanishes to a very good approximation, which yields the 
following approximation for the formation factor:
\begin{align}
  \mathcal{F} \approx \frac{2+\phi_2-\phi_1\zeta_2}{\phi_1(2-\zeta_2)}.
  \label{eq:3ptBound}
\end{align}
Here $\zeta_2\in[0,1]$ is a \textit{three-point microstructural 
parameter}, which is a weighted integral involving the one-, two-, and 
three-point correlation functions $S_1$, $S_2$, and $S_3$.
The high predictive power of Eq.~\eqref{eq:3ptBound} has already been 
validated by excellent agreement with computer simulations of 
$\cal F$ for a variety of ordered and disordered dispersions of spheres 
in a matrix~\cite{torquato_effective_1985, kim_effective_1991, 
robinson_electrical_2005, gillman_third-order_2014, 
gillman_third-order_2015, nguyen_conductivity_2016}.
When $\zeta_2=0$, Eq.~\eqref{eq:3ptBound} reduces to the well-known 
two-point Hashin-Shtrikman lower bound on $\mathcal{F}$ (which is 
optimal for given one- and two-point correlation functions $S_1$ and 
$S_2$)~\cite{hashin_variational_1963, torquato_random_2002}.

\begin{figure}[t]
  \centering
  \includegraphics[width=\linewidth]{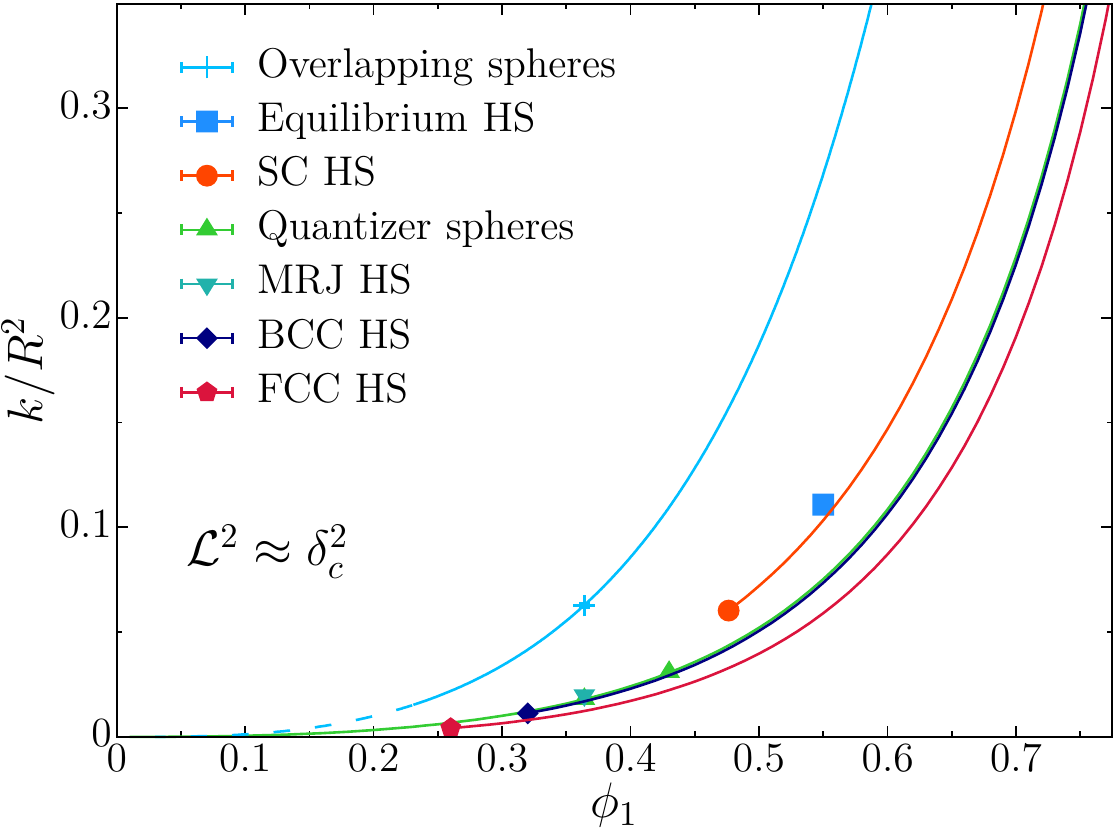}
  \caption{Estimate of permeability as a function of porosity for our 
  hard- and overlapping-sphere models, where $\mathcal{L}^2$ is 
  approximated by $\delta_c^2$.}
  \label{fig:permeability}
\end{figure}

Here we use for our lattice sphere packings the tabulated values of 
$\zeta_2$ up to the maximal packing fractions from Table~20.1 (on 
p.~523) in \citet{torquato_random_2002}, which is based on data from 
\citet{mcphedran_bounds_1981}.
We interpolate the values of $\zeta_2(\phi_2)$ using fourth-order 
polynomials.
For overlapping and equilibrium hard spheres, we use the tabulated values 
of $\zeta_2$ from Table~22.1 (on p.~598) in 
\citet{torquato_random_2002}, which is based on data from 
\citet{torquato_third-order_1985}
and \citet{miller_effective_1990}, respectively.
In these two cases of disordered spheres, an interpolation with 
third-order polynomials was sufficient.
We use the polynomial fit to the equilibrium hard-sphere data also for 
an extrapolation to $\phi_2=0.64$, i.e., to estimate  
$\zeta_2\approx0.148$ for the MRJ sphere packings.
Since no data for $\zeta_2$ is yet available for our quantizer packings, 
we use the Hashin-Shtrikman lower bound in this case.

Figure~\ref{fig:permeability} shows the resulting estimate of the fluid 
permeability $k$ using the approximation by Katz and 
Thompson~\cite{katz_quantitative_1986},
where we choose the empirical proportionality constant between 
$\mathcal{L}$ and $\delta_c$ to be unity, i.e., 
$\mathcal{L}\approx\delta_c$.
The estimate of $k$ is highest for the uncorrelated overlapping spheres 
(among our models and range of porosities); in particular, $k$ is higher 
for the overlapping spheres than for the hard-sphere models (both 
ordered and disordered), which is consistent with the theoretical 
predictions from Ref.~\cite{torquato_predicting_2020}.

\begin{figure}[t]
  \centering
  \includegraphics[width=\linewidth]{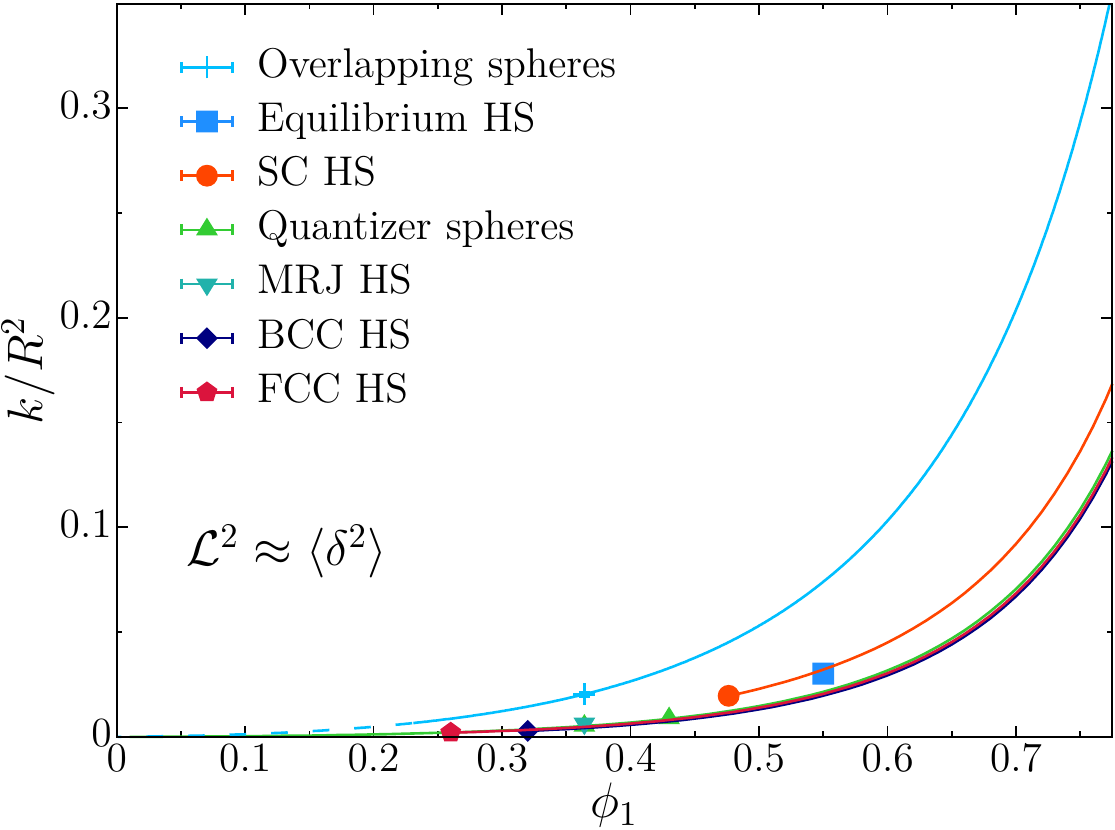}
  \caption{Estimate of permeability as a function of porosity for our 
  hard- and overlapping-sphere models, where $\mathcal{L}^2$ is 
  approximated by $\langle\delta^2\rangle$.}
  \label{fig:permeability-delta_sq}
\end{figure}

Notably, the approximation by $\delta_c$ in Fig.~\ref{fig:permeability} 
provides an inaccurate ranking of the fluid permeability of 
BCC and FCC sphere packings compared to theoretical calculations of 
the fluid permeability~\cite{sangani_slow_1982}.
This inaccuracy is due to the approximation of $\mathcal{L}$ by 
$\delta_c$ rather than the approximation of $\mathcal{F}$, since 
we obtain the same ranking using the Hashin-Shtrikman and three-point 
approximations.
In contrast, the approximation $\mathcal{L}^2 \approx \langle \delta^2 
\rangle$ results in the correct ranking of the fluid permeability $k$ 
for FCC and BCC sphere packings, as shown in 
Fig.~\ref{fig:permeability-delta_sq}.
Moreover, except for the quantizer model that was not studied in 
Ref.~\cite{torquato_predicting_2020}, it was shown that the 
approximation $\mathcal{L}^2 \approx \langle \delta^2 \rangle$ provides 
the correct ranking for all other models  shown in 
Figs.~\ref{fig:permeability} and \ref{fig:permeability-delta_sq}.
Thus, it is reasonable to expect that this approximation would properly 
rank the quantizer model.

\section{Conclusion and outlook}
\label{sec:conclusion}

We have determined the percolation threshold for void percolation around 
sphere configurations for models of both hard and overlapping spheres.
Our examples include the MRJ packings of spheres, equilibrium fluids of 
hard spheres, overlapping spheres, and inherent structures of the 
quantizer energy, as well as ordered lattice packings of hard spheres.

To accurately determine the critical pore radius for our models, we use 
the strict relation to a weighted bond percolation on the Voronoi 
network.
Moreover, we employ the Newman-Ziff algorithm and carefully take 
finite-system size effects into account. We compare our results in 
Table~\ref{tab:critical}
to the second moment of the pore size $\delta$.

We find in Fig.~\ref{fig:cf-delta} a remarkably good correlation 
between $\delta_c^2$ and $\langle \delta^2 \rangle$ across our broad 
spectrum of highly ordered and disordered sphere configurations, 
confirming the suggestion by \citet{torquato_predicting_2020}.
Since $\langle \delta^2 \rangle$ can be easily measured from two- or 
three-dimensional digitized images of heterogeneous materials, this recent approximation 
of $\mathcal{L}^2$ by $\langle \delta^2 \rangle$ allows for a simple yet 
reliable prediction of the permeability $k$.
In fact, we find that, in contrast to the critical pore size $\delta_c$, 
the second moment of the pore size, $\langle \delta^2 \rangle$, predicts 
the correct ranking of $k$ for our models.

Moreover, we observe that the hyperuniform and effectively hyperuniform 
models, like the most hyperuniform BCC sphere packing or the disordered 
MRJ and quantizer packings, tend to have smaller estimates of $k$ than 
the nonhyperuniform overlapping or equilibrium hard spheres.
This again agrees with theoretical arguments from 
\citet{torquato_predicting_2020} that $k$ can be expected to be lower in 
hyperuniform than in nonhyperuniform porous media because the latter 
exhibit a greater variability in the sizes and geometries of the pore 
channels.
Hence, the velocity fields will be generally more uniform throughout the 
pore space for hyperuniform two-phase media compared to their 
nonhyperuniform counterparts.
This is also consistent with the fact that the BCC sphere packings have 
the lowest fluid permeabilities, since the BCC lattice is the structure 
with the lowest value of the hyperuniformity order metric, implying that 
it suppresses large-scale density fluctuations to the greatest 
degree~\cite{torquato_local_2003,torquato_hyperuniform_2018}.
It is interesting to point out that the BCC lattice is also the optimum 
of the covering and quantizer problems~\cite{torquato_reformulation_2010}.
Our results provide additional confirmation for the analysis presented 
in \citet{torquato_predicting_2020} for the aforementioned link between these 
optimization problems, the pore statistics, and fluid 
permeability.

Since the empirical Katz-Thomson formula has already been applied to a broad 
variety of microstructures~\cite{katz_quantitative_1986, 
martys_length_1992, nishiyama_permeability_2017},
a possible direction for future research is to test the 
approximation of $\delta_c^2$ by $\langle \delta^2 \rangle$ for 
polydisperse sphere configurations and more complex particle shapes,
that is, for more general models of porous media as long as the 
pore-space remains well-connected.
This condition is important for the theoretical arguments of the 
approximation of $\mathcal{L}^2$ by $\langle \delta^2 \rangle$.

An important outstanding problem is then to directly determine fluid 
permeabilities from Stokes-flow simulations (as suggested in 
Ref.~\cite{torquato_predicting_2020}).
Another direction for future research is the determination of other 
transport properties besides the permeability, e.g., the effective 
electrical or thermal conductivity of void space (possibly represented 
by the Voronoi network).

\begin{acknowledgments}
  We thank Jaeuk Kim for his samples of equilibrium hard spheres.
  M.~A.~K.~and S.~T.~were supported in part by the Princeton 
  University Innovation Fund for New Ideas in the Natural Sciences and 
  by the Air Force Office of Scientific Research Program on Mechanics of 
  Multifunctional Materials and Microsystems under Award 
  No.~FA9550-18-1-0514.
  M.~A.~K.~also acknowledges funding by the Volkswagenstiftung via the 
  Experiment-Projekt Mecke.
\end{acknowledgments}


%

\end{document}